\documentclass[prb,preprint,showpacs,preprintnumbers,amsmath,amssymb]{revtex4}

\usepackage[dvips]{graphicx}
\usepackage{dcolumn}
\usepackage{bm}
\usepackage{times}
\usepackage{mathptmx}
\usepackage{enumerate}
\usepackage{graphicx}
\usepackage{ulem}

\begin{document}


\title{Electrostatic topology of ferroelectric domains in YMnO$_3$}

\author{Tobias Jungk$^1$}
\author{\'{A}kos Hoffmann$^1$}
\author{Manfred Fiebig$^2$}
\author{Elisabeth Soergel$^1$}
\email{soergel@uni-bonn.de}

\affiliation{$^1$PI, Universit\"{a}t Bonn, Wegelerstra\ss e 8, 53115 Bonn, Germany}
\affiliation{$^2$HISKP, Universit\"{a}t Bonn, Nussallee 14-16, 53115 Bonn, Germany}


\begin{abstract}
Trimerization-polarization domains in ferroelectric hexagonal YMnO$_3$ were resolved in all three
spatial dimensions by piezoresponse force microscopy. Their topology is dominated by electrostatic
effects with a range of 100 unit cells and reflects the unusual electrostatic origin of the
spontaneous polarization. The response of the domains to locally applied electric fields explains
difficulties in transferring YMnO$_3$ into a single-domain state. Our results demonstrate that the
wealth of non-displacive mechanisms driving ferroelectricity that emerged from the research on
multiferroics are a rich source of alternative types of domains and domain-switching phenomena.
\end{abstract}

\maketitle

In materials with a coexistence of magnetic and ferroelectric order, called
multiferroics,\cite{khomskii09,eerenstein06,fiebig05a} the most prominent type of
ferroelectricity, i.e., ferroelectricity of the displacive type found in perovskites like
BaTiO$_3$, is usually avoided.\cite{hill00} Therefore, multiferroics research lead to the
awareness of a wealth of alternative mechanisms driving the emergence of a spontaneous
polarization. This includes ferroelectricity from electron lone pairs, charge order, helical spin
structures, electrostatic effects, and more.\cite{khomskii09,aken04} The unconventional origin of
the spontaneous polarization in these systems should also affect the basic properties of a
ferroelectric such as the distribution and switching behavior of its domains. However, in spite of
the implications of this aspect for technological applications it has not attracted much attention
thus far.

Very recently, the ferroelectric domain structure of hexagonal YMnO$_3$ was
investigated.\cite{choi10} Hexagonal manganites are textbook multiferroics in which the
spontaneous polarization is induced by electrostatic instead of displacive effects. Choi et al.\
found that perpendicular to the spontaneous polarization intriguing intersections of as many as
six ferroelectric domains are common to YMnO$_3$ and related this to the atomic displacement at
the domain wall. In this letter we show that in addition (or alternatively) the unusual domain
topology is a direct consequence of the electrostatic nature of the ferroelectric state in
YMnO$_3$. This is derived from piezoresponse force microscopy (PFM) measurements showing that in
spite of the anisotropic crystal structure kaleidoscopic intersections of domain walls are present
in {\it all three} spatial dimensions. At the intersections electrostatic repulsion leads to a
relative displacement of polarization and trimerization domains that are otherwise rigidly
coupled. PFM tip-poling experiments confirm that the electrostatic discontinuity at the domain
walls controls the spontaneous polarization within a range of about 100 unit cells.

\begin{figure}[h]
\begin{center}
\includegraphics{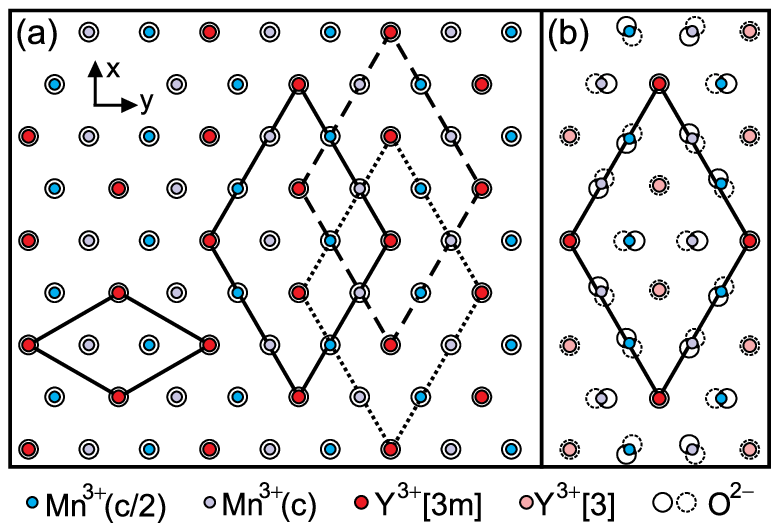}
\end{center}
\caption{\label{fig1} (Color) (a) YMnO$_3$ crystal at $>1270$~K with the unit cell (small diamond)
and three choices for the trimerization at 1270~K (large diamonds). (b) Ferroelectric crystal with
tripled unit cell. Legend --- Mn$^{3+}$ ions: $z$ position in the unit cell; Y$^{3+}$ ions: local
symmetry.}
\end{figure}

In hexagonal YMnO$_3$ the Mn$^{3+}$ ions are found in a rare fivefold coordination with the
O$^{2-}$ ions. Planes of MnO$_5^{3+}$ bipyramids are interspaced with planes of Y$^{3+}$ ions
along the hexagonal $z$ axis. Ferroelectricity emerges in two steps.\cite{lonkai04,fennie05} At
1270~K tilting of the MnO$_5^{3+}$ polyhedra and corrugation of the Y$^{3+}$ layers occurs. At
920~K additional displacement of the MnO$_5^{3+}$ polyhedra induces a spontaneous polarization
$P_z$ with a saturation value of 5.6~$\mu$C/cm$^2$. The transitions are driven by electrostatic
and geometric effects, rather than by the usual changes in chemical bonding associated with
displacive ferroelectric phase transitions as in perovskite oxides. This mechanism permits the
coexistence of magnetism and ferroelectricity so that the compound becomes multiferroic at
cryogenic temperatures.\cite{aken04} As shown in Fig.~\ref{fig1} unit-cell tripling at 1270~K
leads to three trimerization domains ($\alpha$, $\beta$, $\gamma$). The polarization along $+z$ or
$-z$ emerging at 920~K leads to a total of six domains ($\alpha^{\pm}$, $\beta^{\pm}$,
$\gamma^{\pm}$). The trimerization domains are translation domains for which the identification as
$\alpha$, $\beta$, or $\gamma$ is ambiguous. In contrast, the polarization domains are
$180^{\circ}$ domains with a unique assignment of $+$ and $-$. From a variety of experiments
as-grown ferroelectric domains in YMnO$_3$ are known to possess an extension of $\lesssim
1$~$\mu$m.\cite{safrankova67,oleinik75,fiebig02,neacsu09} but a detailed study of their topology
has only been presented by Choi et al.

YMnO$_3$ samples were flux-grown $z$-oriented platelets with a lateral extension of a few mm and a
thickness in the order of 100~$\mu$m.\cite{kim00} Chemical-mechanical polishing with a silica
slurry was applied to the $z$ face and to $x$- and $y$-oriented bars cut from one platelet. The
distribution of ferroelectric domains was measured by PFM which allows us to probe all three
crystallographic directions non-invasively with an impressive sensitivity and high spatial
resolution.\cite{jungk08,jungk09}

\begin{figure}[h]
\begin{center}
\includegraphics{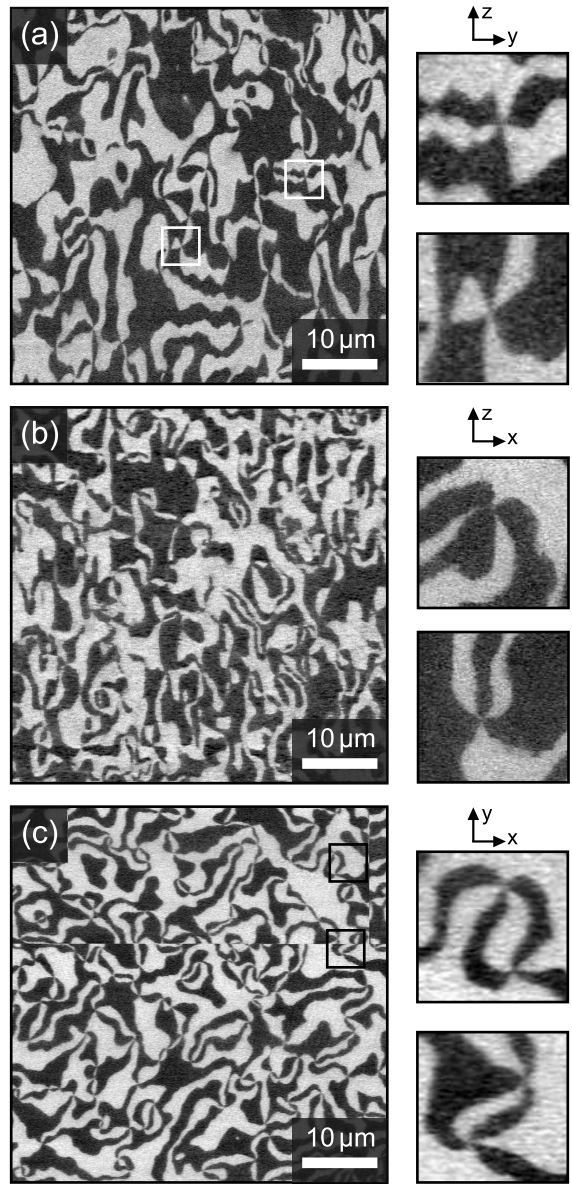}
\end{center}
\caption{\label{fig2} PFM images ($50\times 50$~$\mu$m$^2$) of the $x$, $y$ and $z$ faces of
as-grown YMnO$_3$ crystals from the same batch. Bright and dark areas correspond to ferroelectric
domains with $+P_z$ and $-P_z$, respectively.}
\end{figure}

Figure~\ref{fig2} shows the $x$, $y$, and $z$ face{\bf s} of as-grown YMnO$_3$ crystals from the same
batch under ambient conditions. As expected, all faces reveal domains of $\lesssim 1$~$\mu$m with
two grey levels corresponding to domains with $+P_z$ (bright) or $-P_z$ (dark). However, the
topology of the ferroelectric domains is striking. On the $z$ face of the crystal the same
kaleidoscopic domain structure with meeting points of six domains as in Ref.~\onlinecite{choi10}
is obtained. In Ref.~\onlinecite{choi10} the intersections were assigned to $\alpha^+$, $\beta^-$,
$\gamma^+$, $\alpha^-$, $\beta^+$, $\gamma^-$ (anti-) vortices of domains with a rigid clamping of
trimerization and polarization domain walls. The clamping was attributed to the microscopic
structure at the domain wall: The coexistence of trimerization and polarization walls minimizes
the displacement of the Y$^{3+}$ position with respect to the paraelectric phase which might be
energetically favorable. This argument referred to domain walls in the $xy$ plane so that
kaleidoscopic intersections are expected in this plane only. However, Fig.~\ref{fig2} reveals that
in spite of the highly anisotropic uniaxial structure of the YMnO$_3$ crystal the distribution of
the domains is almost isotropic with only a slight elongation along $z$. In {\it all three}
spatial dimensions the peculiar ferroelectric domain topology with meeting points of six domains
is observed. We therefore conclude that in addition (or alternative) to mechanisms rooting in the
microscopic structure of the YMnO$_3$ unit cell a more general mechanism that is less sensitive to
the microscopic anisotropy must be responsible for the domain structure in Fig.~\ref{fig2}.

\begin{figure*}[ppp]
\begin{center}
\includegraphics{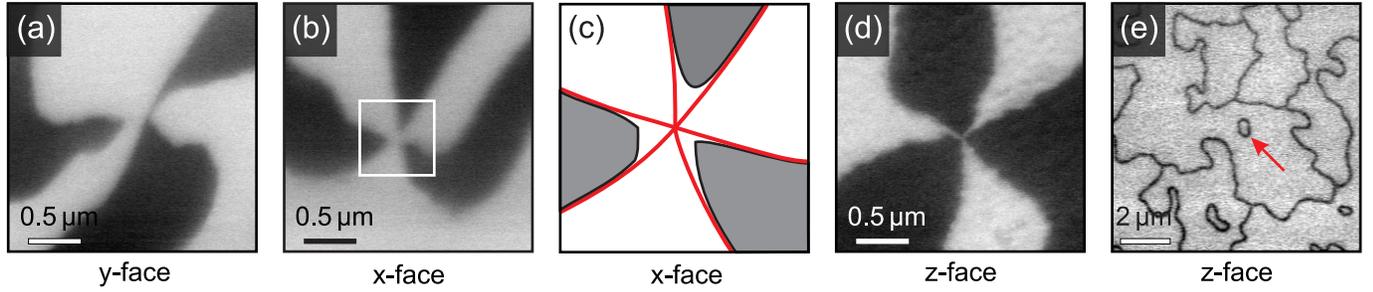}
\end{center}
\caption{\label{fig3} (Color) High-resolution PFM images of the trimerization-polarization
domains. (a, b, d) Exemplary images of the $y$, $x$, and $z$ face as discussed in the text. (c)
Sketch of the domain structure in (b) showing trimerization and polarization walls as straight red
lines and boundaries of grey areas, respectively. (e) Near-single $+P_z$ domain with $-P_z$
domains shrunk to stripes of $\sim 60$~nm. Note that closed-loop domains (arrow) are also
observed.}
\end{figure*}

For elucidating this issue we conducted high-resolution PFM scans of the domain vortex region. The
images in Fig.~\ref{fig3} reveal that the six trimerization-polarization domains do not really
meet in one point. In the majority of cases three equally polarized domains approach one another
up to a distance of about 30~nm on the $z$ face and of about 100~nm on the $x$ and $y$ faces.
Hence, at the center of the corresponding domain vortex the spontaneous polarization is $+P_z$ or
$-P_z$ instead of approaching zero. In a minority of cases two domains of equal polarization are
connected via a thin bridge but separated from the third, equally polarized domain. On the $x$ and
$y$ faces this scenario is met an order of magnitude less often than the former one. For the $z$
face the occurrence of the second scenario is not clear. As in the case of Fig.~\ref{fig2} the
local microscopy of the domain walls cannot be responsible for the topology of domains observed
here. Domain walls in ferroelectrics are at best a few unit cells wide while separations in the
order of 100 unit cells are observed in Fig.~\ref{fig3}. This rather indicates that Coulomb
interactions determine the domain structure at the vortex. The Coulomb force is a central force,
thus acting in all three spatial dimensions and it is a long-range force that can act across many
unit cells.

\begin{figure}[ppp]
\begin{center}
\includegraphics{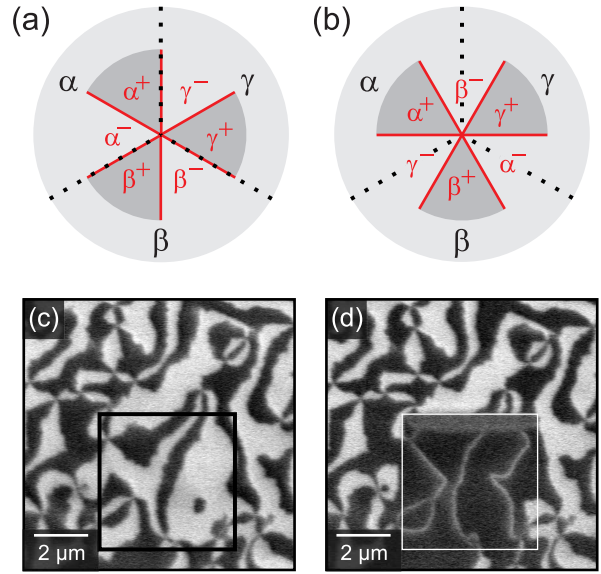}
\end{center}
\caption{\label{fig4} (Color) Structure of trimerization and polarization domains and their
response to an electric field. (a, b) Scenarios 1 and 2 as discussed in the text. $\alpha$,
$\beta$, $\gamma$ and dotted black lines denote the distribution of domains in the trimerized
phase at $>920$~K. $\alpha^{\pm}$, $\beta^{\pm}$, $\gamma^{\pm}$ and straight red lines denote the
distribution of domains in the ferroelectric phase at $<920$~K. (c, d) PFM image ($10\times
10$~$\mu$m$^2$) of a $z$ faced YMnO$_3$ crystal (c) before and (d) after tip-voltage poling.}
\end{figure}

In order to understand the relation of the trimerization and polarization domain walls to the
electrostatic repulsion and vortex structure of the domains we subjected the YMnO$_3$ crystal in
Fig.~\ref{fig4} to an electric poling field. A DC voltage of 50~V was applied via the SFM tip
along the $z$ axis while scanning a square of $5\times 5$~$\mu$m$^2$. In contrast to
Ref.~\onlinecite{choi10} the voltage was applied non-invasively and the same region was imaged
before and after poling. For the domain vortex we have to distinguish between the two scenarios
depicted in Figs.~\ref{fig4}(a) and \ref{fig4}(b): (1) an $\alpha^+$, $\alpha^-$, $\beta^+$,
$\beta^-$, $\gamma^+$, $\gamma^-$ and (2) an $\alpha^+$, $\beta^-$, $\gamma^+$, $\alpha^-$,
$\beta^+$, $\gamma^-$ sequence of domains.\cite{footnote} Scenario (1) involves two types of
boundaries: polarization-trimerization and polarization-only walls. This may occur if the
trimerization walls are so stable that each trimerization domain splits into two polarization
domains. Scenario (2) involves polarization-trimerization walls only. This will occur if the
creation of additional trimerization domains costs less energy than the creation of a
polarization-only wall. Scenario (2) was proposed in Ref.~\onlinecite{choi10} but not explicitly
confirmed there because trimerization and polarization domains were observed in separate
experiments, i.e., by force and electron microscopy, respectively.

According to the scan shown in Fig.~\ref{fig4}(d) {\it all} walls respond to the electric field,
thus indicating that scenario (2) as depicted in Fig.~\ref{fig4}(b) holds. This result was
confirmed by annealing experiments in which a sample was heated to 1150~K for three hours. After
re-cooling all the domain walls, including the trimerization walls, were found to have moved. We
therefore conclude that in YMnO$_3$ trimerization  and polarization walls are rigidly coupled
except in the vortex region where the correlation becomes ``flexible'' with a relative spatial
displacement of trimerization  and polarization walls as shown in Fig.~\ref{fig3}(c).

Note that in contrast to displacive ferroelectrics it is not possible to convert the entire
YMnO$_3$ sample into a single-domain state. Figure~\ref{fig4}(d) shows that the presence of
remanent interstitial domains polarized oppositely to the applied electric field are enforced by
the electrostatic discontinuity at the domain wall. This nicely reflects the electrostatic nature
of the ferroelectric state and explains why in former experiments evidence for a multi-domain
state was still obtained when the coercive field was exceeded by an order of
magnitude.\cite{fiebig02} Quantitative analysis of Fig.~\ref{fig4}(d) reveals a width of $60 \pm
10$~nm for the interstitial domain which matches the electrostatic reach found in Fig.~\ref{fig3}.

Our observations thus reveal the following scenario. Growth-induced trimerization leads to
intersections where one $\alpha$, one $\beta$, and one $\gamma$ domain meet. The electrostatic
discontinuity associated to a trimerization wall controls the ferroelectric polarization up to a
distance of about 100 unit cells. It enforces a reversal of polarization whenever a trimerization
wall is crossed. The simplest arrangement of domains satisfying these requirements involve
kaleidoscopic intersection of six domains as shown in Fig.~\ref{fig4}(b) and
Ref.~\onlinecite{choi10}. At the intersection the electrostatic repulsion between equally
polarized domains is solved by a decoupling of trimerization and polarization domains near the
center of the domain vortex.

In summary the unusual kaleidoscopic topology of ferroelectric trimerization-polarization domains
in multiferroic hexagonal YMnO$_3$ was shown to be a consequence of electrostatic effects which
reflects the unusual electrostatic origin of the spontaneous polarization in this compound. The
discontinuity at the domain walls leads to locally polarizing fields that determine the
spontaneous polarization within a range of 100 unit cells in all three spatial dimensions. At the
intersection of domains this leads to a flexible coupling of trimerization and polarization
domains with correlated, but non-overlapping domain walls. Away from the intersection
trimerization and polarization walls coincide with domains maintaining a width of at least 100
unit cells. Even beyond saturation fields a remanent ferroelectric multi-domain structure is
inherent to the sample which explains former difficulties to transform YMnO$_3$ into a
single-domain state. Our results show that the wealth of non-displacive mechanisms driving
ferroelectricity that emerged from the present intense research activities on multiferroics are a
potential source of alternative domain and polarization phenomena with basic-research as well as
technological implications.

The authors thank the Deutsche Telekom AG and the DFG for financial support and J. F. Scott for
useful advice about multidimensional order parameters.


\newpage


\clearpage

\clearpage

\clearpage

\end{document}